\documentstyle[graphicx,11pt,aaspp4]{article}
\begin{document}


\title{A Spherical Model for ``Starless'' Cores of Magnetic Molecular
Clouds and Dynamical Effects of Dust Grains}

\author{Zhi-Yun Li}

\affil{Astronomy Department, University of Virginia, Charlottesville, 
	VA 22903}

\begin{abstract}

In the current paradigm, isolated low-mass stars form as magnetic
molecular clouds evolve due to ambipolar diffusion. A quantitative
understanding of this process remains incomplete, because of both 
physical and technical complexities. As a step toward a quantitative
theory for star formation, I explore further the evolution of 
magnetic clouds with a simplifying spherical geometry, studied 
first by Safier, McKee \& Stahler. The spherical model
has several desirable features as well as some potentially serious
difficulties. It highlights the pressing need for two-dimensional,
fully dynamic models that treat the coupling between the magnetic 
field and the cloud matter properly. 

The model clouds exhibit substantial inward motion during the 
late stage of core formation, with velocities of order half the 
isothermal sound speed or more over most of the cloud, after 
a central density enhancement of about $10^2$ (say from $10^3$ 
to $10^5$cm$^{-3}$). Such pre-collapse motion may have been 
detected 
recently by Taffala et al. and Williams et al. in the ``starless'' 
core L1544. The motion may also explain why ``starless'' dense 
cores typically last for only a few dynamic times and why they 
have small but significant non-thermal linewidths. The clouds 
that I study have relatively low-mass (of order 10 $M_\odot$) and 
an initial magnetic pressure comparable to the thermal 
pressure. They evolve into a density profile of a flat central 
plateau surrounded by an envelope whose density decreases with 
radius roughly as a power-law, as found previously. The density 
decline in the envelope is significantly steeper than those 
obtained by Mouschovias and collaborators, who considered 
disk-like magnetic clouds whose initial magnetic pressure is 
much larger than the thermal pressure. 

As the cloud density increases above approximately $10^5$cm$^{-3}$, 
dust grains become dynamically important. Depending on their 
size-distribution, dust grains can enhance the coupling 
coefficient between the magnetic field and the neutral cloud
matter by as much as an order of magnitude or even more. 
The enhanced coupling makes it difficult for magnetic flux to 
escape during the advanced, more dynamic phase of the core formation. 
In spherical
geometry, this trapping of magnetic flux leads to an almost homologous 
collapse of a central region with substantial mass. The implied 
extremely rapid assemblage of protostellar mass, with more than 
half a solar-mass material reaching the origin in about $10^3$ 
years or less for a typical set of cloud parameters, poses a serious 
accretion ``luminosity problem''. Magnetic tension in two-dimensional 
models may help alleviate the problem. 
\end{abstract}


\section{Introduction}

The current paradigm for forming isolated, sun-like low-mass stars 
begins with isolated clumps within lightly-ionized molecular clouds. 
The clumps are supported against self-gravity and external pressure by a 
combination of thermal (and perhaps turbulent) pressure and magnetic 
fields. Since 
magnetic forces act on charged particles only, a relative drift 
between the charged species and the 
neutral matter is necessary to transmit
the magnetic support to the predominantly neutral matter. This
relative drift, called ``ambipolar diffusion", reduces the magnetic
flux (and thus magnetic support) gradually in the central region
of the clump, causing the central density to grow quasi-statically. 
When the flux-to-mass ratio drops below certain critical value, 
a ``runaway" collapse ensues, forming a protostar at the center (see 
reviews by Nakano 1984; Shu, Adams \& Lizano 1987; Mouschovias 1994). 
This star formation process was given the picturesque name of 
``gravomagneto-catastrophe" by Shu (1995), in analogy to the 
``gravothermo-catastrophe" discussed by Lynden-Bell \& Wood (1968) 
in the context of stellar cluster evolution (see \S8.2 of Binney and 
Tremaine 1987). The end product of the ``gravomagneto-catastrophe" 
is the birth of a new sun-like star.

The above paradigm provides an attractive framework to synthesize 
a wide range of observations, as reviewed by Shu et al. (1987). 
Many quantitative predictions of the paradigm 
remain to be worked out in details, however. One example is the structure
and kinematics of dense cores at various stages of 
their evolution, for which detailed observational data are 
now becoming available (Myers
1995). Indeed, a recent detection of extended inward motion 
in the ``starless" core  
L1544 (Tafalla et al. 1998) prompts the authors to conclude that
``If this core is in the process of forming stars, our observations
suggest that it is doing so in a manner not contemplated by the 
standard theories of star formation. Our study of L1544 illustrates
how little is still known about the physical conditions that 
precede star formation, and how detailed studies of starless
cores are urgently needed." The standard picture, as outlined
above, may still be able to explain such seemly 
contradictory observations. Of course, we cannot be certain until 
quantitative predictions of the cloud dynamics according to the 
standard picture are made. The technical difficulty
of such a task is daunting: the dynamic range required going
from interstellar clouds to compact protostellar objects is
enormous. This aspect alone would be difficult to handle in itself. 
The inclusion of magnetic field and its subtle coupling to the 
cloud matter makes the task even more challenging.  

Impressive theoretical progresses have been made on the dynamical 
aspect of the star formation problem over the last three 
decades under various simplifying assumptions. We can roughly 
divide previous investigations into three periods. In the first 
period, most of studies were concentrated on nonmagnetic, nonrotating 
isothermal spheres, as exemplified by 
various classes of self-similar solutions found by Larson (1969) and 
Penston (1969), by Shu (1977) and by Hunter (1977). In the second period,
well-ordered magnetic fields were included in the initial equilibrium
configurations of the clouds (Mouschovias 1976) and their evolution 
due to
ambipolar diffusion was followed using quasi-static numerical codes 
(Nakano 1979; Lizano \& Shu 1989). The process of core formation out
of the background can be investigated up to a central density enhancement
of a few hundreds. Beyond such an enhancement the quasi-static assumption 
begins to break down and a dynamic code is needed to follow through the
rest of the core formation process. In the third period, Mouschovias and 
coworkers
have developed dynamic codes that can follow the density increase by
a factor of $10^6$, close to the formation of a central protostar 
(e.g., Basu \& Mouschovias 1994; Ciolek \& Mouschovias 1994). 
These are essentially one dimensional calculations, making good use
of the fact that the cloud matter settles along magnetic
field lines into a disk-like configuration when the plasma-$\beta$
(defined conventionally as the ratio of thermal and magnetic 
pressures) is small  (Fiedler \& Mouschovias 1993). They left 
open, however, the question of the evolution of magnetic clouds
with the magnetic pressure comparable to, or even smaller than, the
thermal pressure. Such clouds are perhaps more sphere-like than
disk-like (see Li \& Shu 1996 for a sequence of cloud configurations
with various degree of magnetization). 

Motivated by the roundish appearance of many dense molecular cores
and as a step towards a quantitative understanding of the star
formation process, I choose to study the evolution of magnetic
clouds in a spherical geometry. Such a simplified geometry 
overcomes the dynamic range problem almost trivially, with a 
Lagrangian code (e.g., Foster \& Chevalier 1993). Magnetic fields 
tend to introduce
anisotropy to the cloud mass distribution (Li \& Shu 1996) and 
strictly speaking two-dimensional (2-D) codes are required. Safier, 
McKee \& Stahler (1997) made the ingenious 
suggestion of driving ambipolar diffusion with magnetic pressure 
gradient alone, which made the reduction to 1-D possible
while retaining the essence of the dynamical problem. It is borne out
by subsequent numerical calculations of Li (1998) who, using a 
Eulerian code following Mouschovias \& Morton (1991), showed that the
evolution of magnetized spherical clouds 
is broadly similar to those studied by Mouschovias and collaborators.
As with their thin-disk simplification in the case of relatively 
strong magnetic fields, 
the spherical geometry allows one to investigate additional 
physical processes that are important to the star formation 
process but are currently difficult to be incorporated into genuinely
2-D, dynamic codes, such as that of Fielder \& Mouschovias (1993). 
The work of Li (1998) provides a starting point for the present
investigation. 

There are two major improvements of the present study over that of Li 
(1998): 1) a Lagrangian instead of Eulerian method
implemented to enhance the dynamic range, taking full advantage of the 
simplified geometry;  
and 2) detailed calculations of number densities of charged species 
(including dust grains) from cosmic ray ionization. The use of
Lagrangian coordinates was the original approach of Safier et
al. (1997). They found the coordinates convenient for obtaining 
analytic solutions for the quasi-static epoch of the cloud evolution.  
These analytic solutions were obtained by omitting the ion velocity and 
the thermal pressure. The inclusion of both quantities in the 
dynamic as well as the quasistatic 
epochs of the cloud evolution necessitates a numerical 
treatment of the problem, as done in this paper. The second 
improvement is particularly important because the number densities
of the charged species determine the coupling coefficient between 
magnetic fields and the cloud matter, which controls the cloud
evolution. The field-matter coupling is where the dynamical 
effects of dust grains come
in. As shown by Nakano and coworkers (e.g., Nakano 1984; Nakano
\& Umebayashi 1986; Nishi, Nakano \& Umebayashi 1991) and others, dust 
grains play a crucial role in the magnetic coupling. Not only can 
dust grains be charged and tied to magnetic fields themselves, they 
substantially modify the number 
densities of other charged species -- ions and electrons -- as well. 
Additional improvements, such as the inclusion of time-dependent 
chemistry (necessary for understanding, for example, the chemical 
differentiation observed by Kuiper, Langer \& Velusamy 1996 in
L1498) and its effects on the cloud dynamics, will be treated 
in the future. 

To gauge the dynamical effects of dust grains, I shall first consider a 
reference model in which the magnetic coupling coefficient is fixed
at a canonical value (approximately valid in general at densities below
about $10^5$cm$^{-3}$ and over a wider density range for certain
grain-size distributions; see \S 5.2), and then an improved 
model in which the
coupling coefficient is evaluated self-consistently from ionization 
equilibrium, taking into account the size distribution of dust grains.
I shall discuss the model predictions, their limitations, and their 
implications for the structure and kinematics of dense molecular cores 
(especially those ``starless" ones that are yet to harbor embedded 
stellar sources) as well as the protostellar mass growth rate. 

\section{Governing Equations for Cloud Evolution}

The governing equations for the evolution of a spherical, magnetized
cloud in an Eulerian form are given by Li (1998; his equations 6-10). 
Here I rewrite these equations into a Lagrangian form as
\begin{equation}
{\partial r\over \partial M}={1\over 4\pi \rho r^2},
\end{equation}
\begin{equation}
{\partial V\over\partial t}=-{G M\over r^2}-4\pi r^2{\partial\over
\partial M}\left(\rho a^2 +{B^2\over 8\pi}\right),
\end{equation}
\begin{equation}
{\partial\over\partial t}\left({B\over 4\pi\rho r}\right)=
{\partial\over\partial M}\left[ B r(V-V_{_B})\right],
\end{equation}
\begin{equation}
{\partial r\over \partial t}=V,
\end{equation}
where $V$ and $V_{_B}$ are the flow speed of the predominantly neutral 
cloud matter and the speed of the magnetic field lines respectively, $r$ 
the spherical radius, $M$ the mass enclosed within a sphere, $\rho$ 
the mass density, and $B$ the magnetic field strength. A constant 
isothermal sound speed of $a$ is assumed for the cloud. If ions are
the main provider of the magnetic coupling and are well-tied to the field 
lines, then $V_B=V_i$ (ion velocity), and the coupling of the magnetic
field to the neutral matter is determined by the ion-neutral collision 
timescale
\begin{equation}
t_{ni}={V_i-V\over f_B/\rho_{H_2}} = {1\over \gamma \rho_i},
\end{equation}
where $f_B=-\partial (B^2/8\pi)/\partial r$ is the magnetic force,
$\rho_{H_2}=\rho/1.4$ the mass density of hydrogen molecules (We 
neglect the collision of ions with helium atoms following Mouschovias
\& Morton 1990), $\rho_i$ the ion density and $\gamma$ a coupling 
constant. In the more general case where magnetic fields are not
necessarily tied to ions, we can define a magnetic field-neutral 
coupling timescale
\begin{equation}
t_{nB}={V_B-V\over f_B/\rho_{H_2}} 
\end{equation}
analogous to $t_{ni}$ in equation (5). It is the timescale that it 
would take the magnetic force to accelerate hydrogen molecules 
to the drift velocity between the magnetic field and the neutrals. The
evaluation of this timescale is nontrivial, and needs to be done  
numerically (see \S~4). 

To cast the above equations into a nondimensional form ready for a
numerical attack, we follow Li (1998) by scaling the density and the
field strength by their initial values at the cloud center, $\rho_c$ 
and $B_c$, the time by the initial central free-fall timescale
\begin{equation}
t_{ff,c}={1\over (4\pi G\rho_c)^{1/2}},
\end{equation}
and the speed by the initial central Alfv\'en speed 
\begin{equation}
V_{_A,c}={B_c\over (4\pi\rho_c)^{1/2}}.
\end{equation}
The natural scales for the radius and the mass are then $r_c=V_{_A,c}t_{ff,c}$
and $M_c=4\pi\rho_c r_c^3$ respectively. With these scalings, we can
rewrite equations (1)-(4) into
\begin{equation}
{\partial \xi\over \partial m}={1\over\hat{\rho}\xi^2},
\end{equation}
\begin{equation}
{\partial u\over\partial\tau}=-{m\over \xi^2}-\xi^2{\partial\over\partial
	m}\left({\hat{\rho}\over 2\alpha_c}+{b^2\over 2}\right),
\end{equation}
\begin{equation}
{\partial\over\partial\tau} \left({b\over \hat{\rho}\xi}\right)={\partial
	\over\partial m}\left({1.4\over\nu_{ff}}{b^2\xi^3\over 
	\hat{\rho}^{1/2}}{\partial b\over\partial m}\right),
\end{equation}
\begin{equation}
u={\partial \xi\over \partial\tau},
\end{equation}
where the dimensionless variables are
\begin{equation}
\xi={r\over r_c};\ m={M\over M_c};\ \hat{\rho}={\rho\over \rho_c};\
u={V\over V_{_A,c}};\ \tau={t\over t_{ff,c}};\ b={B\over B_c}.
\end{equation}
Note that equation (11), which governs the evolution of the 
flux-to-mass ratio, is obtained by combining equation (3)
and the mass continuity equation. 

There are two dimensionless constants that appear in the above equations.
One of them, $\alpha_c$, denotes the initial ratio of the magnetic pressure 
and the thermal pressure at the cloud center, $B_c^2/(8\pi\rho_c 
a^2)$. The other is the magnetic coupling coefficient $\nu_{ff}$, 
defined as the ratio of the local free-fall timescale  
$t_{ff}=(4\pi G\rho)^{-1/2}$ and the magnetic field-neutral
coupling timescale $t_{nB}$ defined in equation (6). That is, 
\begin{equation}
\nu_{ff}\equiv {t_{ff}\over t_{nB}}={1.4\over (4\pi 
	G)^{1/2}}{1\over\rho^{3/2} (V-V_B)}{\partial\over
	\partial r}\left({B^2\over 8\pi}\right). 
\end{equation}
In the special case where ions are
the main provider of the magnetic coupling and are well-tied 
to the field lines, the expression for $\nu_{ff}$ reduces to
the familiar form 
$\nu_{ff}=\gamma\rho_i/(4\pi G\rho)^{1/2}$. 
It is called ``the collapse
retardation factor'' by Ciolek \& Mouschovias (1994). In the canonical 
case where
$\rho_i=C\rho^{1/2}$, the coupling parameter $\nu_{ff}$ is a constant
independent of the cloud density and field strength, 
with a nominal value of about 10 (Galli \& Shu 1993; 
Ciolek \& Mouschovias 1994). We shall use 
the case with $\nu_{ff}=10$ as a reference to gauge the effects
of a detailed treatment of the magnetic field-matter coupling, where
the coefficient is computed self-consistently for each given pair 
of density and field strength from ionization equilibrium. 

\section{Reference Model with Constant Magnetic Coupling Coefficient}

\subsection{Initial and Boundary Conditions}

Dense NH$_3$ cores of molecular clouds have typical number densities of a 
few times $10^4$cm$^{-3}$ (Myers 1995). They are believed to be formed out 
of clumps of lower densities, such as those probed by $^{13}$CO which 
have typical number densities of a few times $10^3$ cm$^{-3}$. As in 
previous studies of molecular
cloud evolution (e.g., Lizano \& Shu 1989; Ciolek 
\& Mouschovias 1994; Basu \& Mouschovias 1994), I adopt the lower-density
clumps as the starting point for our calculation, even though how they 
are formed in the first place remains an open question. One suggestion
is that they are formed out of magnetized, turbulent medium of an even
lower density (Gammie \& Ostriker 1996). For simplicity, I shall ignore 
the turbulent pressure and assume that the magnetic pressure is initially
equal to the thermal pressure everywhere in the cloud. It demands that
$\alpha_c=1$, where $\alpha_c$ is the inverse of the conventional 
plasma-$\beta$
parameter at the cloud center. If $\alpha_c$ is much smaller than
unity, then the magnetic field may not play a significant role in
the cloud dynamics. On the other hand, if $\alpha_c$ is much larger
than unity, then the cloud would probably appear more disk-like
than sphere-like, invalidating our simplification of geometry. In any
case, with $\alpha_c=1$, I find an overall mass-to-flux ratio for
the cloud not far from the critical value of $1/(2\pi G^{1/2})$ 
(see Figure~1f below), consistent with 
currently available Zeeman observations of magnetic fields in dense 
clouds (e.g., Crutcher 1999).  

I shall limit my discussion to relatively low-mass clumps of order ten 
solar masses that form only one dense core at a time. Many isolated, 
star-forming Bok globules seem to fall into this category (Clemens 
\& Barvainis 1988). The clump is idealized as a magnetized sphere, whose
mass is to be compared with the (thermal) Bonnor-Ebert mass of a 
pressure-confined 
sphere. If the clump has a mass less than the Bonnor-Ebert mass, it will 
evolve toward a stable configuration with uniform magnetic field and
become a ``failed core'' in the terminology of Lizano \& Shu (1989). If,
on the other hand, the clump has a mass greater than the Bonnor-Ebert mass,
it is destined to collapse and form stars as magnetic support weakens due 
to ambipolar diffusion. We are interested in the latter case only. For 
definiteness, I choose a clump with a total mass of $10$ M$_\odot$, a
constant temperature of $10$ K, and an initial molecular hydrogen number
density of $n_{H_2,c}=10^3$cm$^{-3}$ at the center. Together with 
$\alpha_c=1$, they yield a static initial cloud configuration with 
the following dimensional units for dimensionless quantities
used in the direct numerical calculations: $\rho_c=4.68\times 10^{-21}$
cm$^{-3}$ for mass density (including $10\%$ of He by number), 
$t_{ff,c}=0.266$ km s$^{-1}$ for velocity ($\sqrt{2}$ times the isothermal
sound speed $a=0.188$ km s$^{-1}$),
$r_c=0.138$ pc for radius, $M_c=2.26$ M$_\odot$ for mass, and $B_c=
6.45\ \mu$G for field strength. Dimensional units for other choices
of cloud parameters can be obtained easily from formulae in the
last section. It turns out that this particular cloud has a total mass
about 1.8 times the Bonnor-Ebert mass and a center-to-edge 
density contrast of about 2.8. 

The cloud would remain in a static equilibrium indefinitely were it not 
for ambipolar diffusion (e.g., Ciolek \& Mouschovias 1994). We turn
on ambipolar diffusion at the time t=0, and follow the cloud evolution
numerically. The cloud evolution is subjected to the following boundary 
conditions at all
times: at the origin, we impose the usual reflection symmetry, which 
demands that the spatial derivatives of the density and the
field strength vanish and that the flow speed be zero. The reflection
symmetry is valid as long as the central density remains finite,
since the thermal pressure should be able to erase any cusp over 
a Jeans lengthscale around the origin. At the outer edge of the cloud, 
we assume a free pressure boundary across which thermal pressure and 
magnetic pressure are balanced separately, as in Mouschovias \& Morton 
(1991). It implies that the density and the field strength at the edge 
are fixed at their initial values at all times. 

\subsection{Numerical Method} 

For each time step, the cloud evolution is carried out numerically in two 
parts: (1) a hydrodynamic part and (2) a magnetic field diffusion 
part. I use a Lagrangian method to treat the hydrodynamics. The method 
is well known and is documented in details in, e.g., Chapter 4 of the 
textbook ``Numerical Modeling in Applied Physics and Astrophysics" by 
Bowers and Wilson (1991). It is well-suited for one dimensional problems
such as the one at hand. Briefly, one first divides the cloud mass into 
a number of zones and
applies the momentum equation (10) at the zone boundaries to update  
the cloud velocity. It is followed by solving equation (12) for a
new set of zone boundary positions. Conservation of mass in each zone
then yields a updated density from equation (9). This finishes the
first part of the calculation. For the second part, one needs to solve 
the diffusion equation (11) for the magnetic field distribution. 
An efficient technique for such a task is the method of Gaussian 
elimination. Interested readers can find detailed discussions of this
method in \S~6.4 of Bowers and Wilson (1991). The whole procedure is 
repeated for each time step of the cloud evolution. 

\subsection{Main Features of the Cloud Evolution}

The cloud is initially in a hydromagnetic equilibrium. At time $t=0$, 
ambipolar diffusion is turned on. The subsequent cloud evolution 
is followed numerically. In column 2 of Table~1, I list the the time 
intervals it takes the cloud to increase its central number density 
by eight successive factors of 10 in the reference model, from $10^3$
to $10^{11}$ cm$^{-3}$. It is clear that the cloud evolves 
at an increasingly fast pace, as found by many previous studies. The 
density formally reaches infinity in a finite amount of time, forming 
a compact stellar system at the center. The increase in the cloud 
density is usually associated with the core formation phase of star 
formation, where the ``core'' is probed observationally by high
density molecular tracers, such as NH$_3$ and CS (e.g., Myers 1995). Various
model properties of the ``core'' are displayed in Figure~1, when the
central density reaches $10^3$ (initial), $10^4$, ..., $10^{11}$cm
$^{-3}$. From the density and magnetic field distributions in panels 
(a) and (b), we find that the cloud always maintains a plateau plus 
envelope structure. The density and field strength are more or less 
uniform in the central plateau region but decrease with radius roughly as 
a power-law in the envelope. One notable feature is that the power 
index for the density, shown in panel (c), is close to -2.3 in the
bulk of the envelope, which is somewhat steeper than -2, the index for
a singular isothermal sphere. Panels (a)-(c) specify the structure of
the cloud. The cloud kinematics is shown in panel (d), where the
contraction speed is plotted against radius. Notice that, in this
idealized spherical model, most of the cloud is contracting at a speed
of order 0.1 km s$^{-1}$ (i.e., about half of the isothermal sound speed
$a=0.188$ km s$^{-1}$) 
or higher when the central number density
exceeds about $10^5$cm$^{-3}$. Such extended pre-collapse
motion may have interesting implications
for the kinematics of ``starless'' cores (to be discussed in \S~6.1). The
extent of cloud contraction can also be seen from panel (e), where
the contraction speed is plotted against mass. The maximum infall 
speed occurs near the 
plateau-envelope boundary, with the plateau contracting ``outside-in''
(i.e., the highest speed at the largest radius) and the envelope 
``inside-out''. All these structural changes and kinematics are resulted
from a redistribution of magnetic fields due to field diffusion
(Nakano 1979; Mouschovias 1994). The
redistribution is
shown most clearly in the distribution of the mass-to-flux ratio, plotted
in panel (f). The ratio increases steadily above the critical value of
unity in most part of the cloud, except in the plateau region. It
indicates that the evolution of the plateau region is controlled to a
large extent by magnetic diffusion. 

\section{Improved Model with Detailed Treatment of Magnetic Coupling} 

\subsection{Ionization Calculation of Nishi, Nakano \& Umebayashi}

Within star-forming molecular clouds shielded from external ultraviolet 
radiation, ionization is mainly due to cosmic rays. The ionization 
level in such predominantly neutral clouds have been investigated 
by several authors, e.g., Elmegreen (1979), Draine \& Sutin (1987), and 
especially Nakano and collaborators (see Nakano 1984 for a review). 
These latter authors developed a simplified reaction scheme which involves 
the following set of charged species: electron (e$^{-}$), H$^+$, He$^+$, 
C$^+$, H$_3^+$, molecular ions (m$^+$; except H$_3^+$), metal ions 
(M$^+$), and charged dust grains. The reactions are initiated through 
the ionization of H$_2$ and He by cosmic rays and radioactive elements. 
Interested readers should consult \S2.1 of Umebayashi \& Nakano (1990) 
for details. 

One of the main uncertainties in calculating cloud ionization level is the size
distribution of dust grains. Dust grains have two opposing effects 
on the coupling between magnetic 
field and the cloud matter. On one hand, small grains have large
total surface area which allows efficient recombination of ions and
electrons on grain surfaces, leading to a weakening of the 
coupling. On the other hand, charged grains (especially small ones)
are themselves tied to the magnetic field, providing an additional 
coupling agent besides ions and electrons. Detailed calculations 
are needed to determine which effect dominates. Nishi et al. (1991;
see their \S~2 for details) 
considered four possible grain-size distributions: (1) the standard 
distribution of Mathis, Rumpl \& Nordsieck (1977; hereafter MRN) with
\begin{equation}
{dn_{gr}\over da}=A n_H a^{-3.5}
\end{equation}
between $50\AA<a<2500\AA$, where $n_{gr}$ is the number density of
grains and the coefficient $A\approx 1.5\times 10^{-25}$ cm$^{2.5}$; 
(2) the MRN distribution with ice mantles 
($90\AA<a<4500\AA$); (3) the extended MRN distribution without ice
mantles ($3\AA<a<2500\AA$), and (4) the standard MRN distribution plus 
an additional population of grains with a size of $a=4\AA$. The 
smallest dust grains in models (3) and (4) were proposed to explain infrared 
observations
of mainly diffuse clouds and reflection nebulae (e.g., Leger \& Puget 
1984). Direct observational evidences for their existence in dense
molecular cloud cores are still lacking (Boulanger et al. 1998). 
For this reason, we shall concentrate on the cloud evolution 
with the standard MRN distribution, i.e., model (1). The case with 
ice-coated grain distribution of model (2) is qualitatively similar 
(see \S~5). 

\subsection{Evaluation of the Magnetic Coupling Coefficient}

As discussed in the introduction, the cloud evolution is initiated 
and controlled by the diffusion of magnetic fields relative to 
the neutral matter. Depending on the number densities of the 
charged species and the field strength, the diffusion falls into two 
conceptually distinctive regimes: the ambipolar diffusion regime 
and the Ohmic dissipation 
regime (Nakano 1984). A unified formulation is provided by Nakano \& 
Umebayashi (1986), who found analytical expressions for the relative 
drift velocity of field lines with respect to neutral matter under the 
action of arbitrary magnetic forces. Magnetic flux enclosed within 
a circle that moves with the drift velocity relative to the neutrals 
is conserved, no matter which of the two field dissipation processes 
dominates. It is the formulation that I shall adopt. It represents a
significant improvement over previous dynamical calculations, 
and will become especially important in the high density regions 
during the late stages of the cloud evolution. I ignore the momentum 
coupling of the neutral grains to the charged grains due to inelastic
charge capture, whose effect has been shown to be minor (Ciolek
\& Mouschovias 1994; Nakano 1984). 

According to Nakano \& Umebayashi (1986; see also Nishi et al. 1991), the 
cross-field velocity of field lines relative to the neutral matter is
given by 
\begin{equation}
V_B-V={A_1\over A}{1\over 4\pi}\vert {\bf (\nabla\times B)}\times 
{\bf B}\vert,
\end{equation}
with 
\begin{equation}
A_1=\Sigma_\nu{\rho_\nu\omega_\nu^2\over\tau_\nu\Omega_\nu^2},\ 
A_2=\Sigma_\nu{\rho_\nu\omega_\nu\over\tau_\nu^2\Omega_\nu^2},\ 
A=A_1^2+A_2^2,\
\Omega_\nu^2={1\over \tau_\nu^2}+\omega_\nu^2,
\end{equation}
where $\rho_\nu$, $\tau_\nu$ and $\omega_\nu=q_\nu eB/(m_\nu c)$ are,
respectively, the density, the stopping time in a sea of neutrals
and the cyclotron frequency of a charged particle $\nu$ with mass
$m_\nu$ and charge $q_\nu e$. From equations (13) and (16), we can 
obtain the following expression for the dimensionless magnetic 
coupling coefficient 
\begin{equation}
\nu_{ff}={1.4\over (4\pi G)^{1/2}}{A\over \rho^{3/2}A_1},
\end{equation}
which controls the field diffusion. 
It depends on the cloud density $\rho$ as well as the number densities 
of all charged species and the field 
strength through $\rho_\nu$ and the cyclotron frequency $\omega_\nu$
in the quantities $A$ and $A_1$. For a given ionization rate by
cosmic rays and radioactive elements,
I compute the equilibrium charge densities as a function of the cloud 
density for the standard MRN grain-size distribution using Nishi's 
ionization code. These charge distributions are shown in Figure~1
of Nishi et al. (1991). They are obtained by dividing the grain
size distribution into 20 bins of equal width in the logarithmic space.
The steady-state rate equations for all species, including dust
grains, are then solved iteratively. 

With the charge density $\rho_\nu$ thus determined, it is easy to 
compute the coupling coefficient $\nu_{ff}$ as a function of 
density and field strength from equations (17) and (18). The result is
shown in Figure~2, where $\nu_{ff}$ is plotted against the
number density of molecular hydrogen for various values of the
magnetic field strength ranging from $1\ \mu$G to $100$ G (with an
increment factor of $10^{0.1}$), assuming a
canonical cosmic ray ionization rate of $10^{-17}$s$^{-1}$ and 
an ionization rate of $6.9\times 10^{-23}$s$^{-1}$ from radioactive
elements (Umebayashi \& Nakano 1990).   

There are several interesting features in Figure~2: (1) in the low
density region (below about $10^5$cm$^{-3}$), the 
coupling coefficient is rather insensitive to either 
density or field strength. It has a value close to the canonical
value of $\nu_{ff}=10$; (2) as density increases, $\nu_{ff}$ can
differ from its canonical value by a large factor. For relatively
weak magnetic
fields, the coupling parameter drops quickly below unity, making
the field essentially decoupled from the cloud matter. For relatively
strong
fields, the coupling parameter can exceed its canonical value by
a factor of 10 or more. Such a strong coupling has important 
consequences for the cloud dynamics, as we shall see shortly; and
(3) at large enough densities, decoupling will occur for any reasonable 
value of field strength. The implication of such a decoupling for the
``magnetic flux problem'' of star formation is discussed by Li \& McKee
(1996) and Ciolek \& K\"onigl (1998). Therefore, the magnetic coupling 
coefficient could be highly variable. This is why it must be 
evaluated self-consistently. Because of the improvement in the treating 
the magnetic coupling, we call the present model the ``improved'' model.
The improvement mostly affects the cloud evolution in the high
density region, beyond about $10^5$cm$^{-3}$. 

\subsection{Cloud Evolution and Comparison to the Reference Model}

With the coupling coefficient $\nu_{ff}$ known as a function of density 
and field strength, one can repeat the cloud evolution calculation 
by integrating equations (8)-(11) simultaneously. I adopt the same 
initial and boundary conditions as in the reference model of \S~3. The 
only difference is that the previously assumed constant coefficient 
$\nu_{ff}=10$ is now replaced by a self-consistently computed function 
$\nu_{ff}(n_{H_2},B)$, shown in Figure~2. The main properties of the 
cloud evolution in this improved model are displayed in Table~1 and 
Figure~3. In Table~1 (column 3), I list the time intervals it takes the
cloud to increase its central number density by eight successive factors 
of 10, from $n_{H_2}=10^3$ to $10^{11}$cm$^{-3}$. These time intervals
are quite comparable to their counterparts in the reference case. The
slightly smaller time intervals at the early times of the cloud 
evolution are due to the fact that the magnetic coupling coefficient 
at low densities is somewhat smaller than the canonical value of 
$\nu_{ff}=10$ used in the reference model (see Figure~2). At higher
densities, the time intervals become slightly longer than those of 
the reference model, because of better magnetic coupling due to
dust grains (see \S~5.1). 

The differences between the improved model and the reference model are
more apparent in the structure and kinematics of the cloud. From 
Figure~3, we find that, while there is still a 
plateau-envelope profile for the density (panel a) and field strength 
(panel b) in the improved model, the distributions in the envelope 
deviate considerably from a single power-law. This behavior 
is seen most clearly in panel (c), where 
the power-index for density distribution is plotted as a function
of radius. A dramatic steepening of the density profile near the 
plateau-envelope boundary is evident. In panel (d), I display the 
velocity profile of the cloud, for both neutrals (solid lines) and 
molecular ions (represented by HCO$^+$; dotted lines). As in the 
reference model (Figure~1d), 
substantial contraction (faster than, say, about half an isothermal 
sound speed) 
is obtained in the bulk of the cloud after the central density goes
beyond about $10^5$cm$^{-3}$. During the later 
times of the cloud evolution, the maximum infall speed can be substantially 
larger in the improved model than in the reference model, by a factor 
of two or more. At the end of the run, 
it reaches a value of $1.55$ km s$^{-1}$, more than eight times the isothermal
sound speed. At the radius where the maximum infall occurs, we have 
a number density $n_{H_2}=3.43\times 10^8$cm$^{-3}$ and a
field strength $B=7.66$ mG, which yield a local Alfv{\'e}n speed
of $V_{_A}=(B^2/4\pi\rho)^{1/2}=0.54$ km s$^{-1}$, almost three
times the sound speed. The corresponding Alfv{\'e}n Mach number is 
therefore $M_{_A}=2.87$, which is not far from the Mach number 
of 3.3 obtained in the self-similar solution of Larson-Penston
for the collapse of nonmagnetic singular isothermal spheres (see
also Basu 1998). Note that, despite the super-Alfv{\'e}nic 
speed, shocks will not form during this phase of the cloud evolution
because individual fluid parcels in the super-Alfv{\'e}nic inflow 
are not decelerated. 

The most striking feature of the improved case is shown in panel
(e) of Figure~3, where the infall speed is plotted against the
mass in solar mass units: a mass of 0.62 M$_\odot$ in the central
region of the cloud, enough to 
build a typical low-mass star, appears to collapse more or less 
homologously at the end of the run. Note also that the mass-to-flux 
ratio, shown in panel (f), is well below that of the reference 
model in the inner part of the cloud. In the next section, we shall
trace these differences to the dynamical effects of dust grains.

\section{Dynamical Effects of Dust Grains}

\subsection{Effects of Dust Grains on Magnetic Coupling}

The only input quantity that is different between the reference model
and the improved model is the 
coupling coefficient between magnetic fields and the cloud matter. In 
the former case, the coefficient is given as a constant. In the latter
case, it is 
calculated self-consistently during the cloud evolution. This difference
must be responsible for the differences in the cloud properties outlined
in the last section. In Figure~4a, we show the distribution of the
self-consistently calculated coupling coefficient 
$\nu_{ff}$ as a function of the cloud number 
density at the same nine times as in Figure~3. Note that the coupling
coefficient is within a factor of two of the canonical value at
low densities with $n_{H_2} < 10^5$cm$^{-3}$. It peaks between
$n_{H_2}=10^7$cm$^{-3}$ and $10^8$cm$^{-3}$, with a peak value of 
almost an order of magnitude larger than the canonical value. At 
still higher densities, the coupling coefficient decreases steadily,
dropping below the canonical value at a number density between 
$10^{10}$ and $10^{11}$cm$^{-3}$. The peak in the distribution of 
the magnetic coupling coefficient results from dust grains. 

Naively, one may not expect dust grains to dominate the magnetic coupling 
since they tend to be more loosely tied to magnetic fields than ions
and they take longer to stop by collision with neutrals than ions (i.e., 
less efficient in momentum exchange). However, both tendencies could 
be compensated by the fact that dust grains are much heavier than ions, 
and thus provide more inertia per unit charge. It is for this same
reason that ions usually contribute more to the magnetic coupling 
than electrons. These opposing effects show up in the expression for
the coupling coefficient, equation (18), written as a sum over 
contributions from individual charged species
\begin{equation}
\nu_{ff}={1.4\over (4\pi G)^{1/2}\rho^{3/2}}{A\over A_1}
	=\Sigma_\nu\left[{1.4\over (4\pi G)^{1/2}\rho^{3/2}}
	\left( {\rho_\nu\omega_\nu^2
	\over\tau_\nu\Omega_\nu^2} + {A_2\over A_1}{\rho_\nu\omega_\nu
	\over\tau_\nu^2\Omega_\nu^2}\right)\right] 
\end{equation}
where the definitions of $A_1$ and $A_2$ have been used. In Figure~4b, 
I plot the contributions from electrons,
all kinds of ions, all dust grains as well as dust grains in each of
the 20 size bins as a function of the cloud density at
the end of the run in the improved model. It is apparent that the
dust grains dominate the coupling at most of the density range except 
near the low-density end ($n_{H_2} < 10^4$cm$^{-3}$), where ions dominate. 
Contribution from electrons is not significant anywhere. Furthermore, small 
grains contribute more to the magnetic coupling than large grains, as 
expected. 

The enhancement of the magnetic coupling coefficient due to dust grains,
especially in the intermediate density region between $n_{H_2}=10^7$ to $10^8$
cm$^{-3}$, makes it difficult for magnetic flux to leave the central 
region of the cloud. The slow leakage of magnetic flux is the reason 
why the central mass-to-flux ratio changes relatively little at the 
late times (see Figure~3f) of the cloud evolution. It also explains
the creation of the more or less homologously contracting ``core'' seen 
in Figure~3e. As the density 
of the core $\rho_0$ increases, its size shrinks approximately as $r_0
\propto\rho_0^{-1/3}$ since little mass is leaving the core. It leads 
to an increase of the field strength approximately as $B_0\propto 
\rho_0 r_0\propto \rho_0^{2/3}$ because of near flux-freezing in
a spherical geometry. As a 
result, the magnetic pressure in the core should scale with the density 
roughly as $P_B\propto B_0^2\propto \rho_0^{4/3}$ (see also Safier 
et al. 1997). It dominates the thermal 
pressure at high densities since the thermal pressure increases more 
slowly with density ($\propto\rho$ for an isothermal cloud). Therefore,
we have a core with an effective adiabatic index of approximately 4/3.
It is well known from  studies of stellar core collapse (Goldreich \&
Weber 1980; Yahil 1983) that cores with such an equation of state tend
to contract homologously. Such a contraction may have important 
implications for the protostellar mass accretion rate (see the next 
section). 

The existence of a nearly homologously contracting core explains the 
differences in the distributions of
density and field strength in the envelope between the improved
and the reference models. Since little mass and 
magnetic flux leak out of the central region into the inner part of the 
envelope in the improved model as compared to the reference model, 
the central density and field strength are higher for the same 
degree of contraction. This
in turn creates a steeper gradient just outside the core than further
out in the envelope, as shown in Figures~3a through 3d. 

I should point out that the homologously contracting core found in the
spherical geometry may disappear in the more realistic axisymmetric 
geometry, even when the magnetic pressure is smaller than the thermal 
pressure (and thus a roundish cloud) to begin with. The enhanced 
magnetic coupling due to dust grains traps magnetic flux, leading to
a more rapid increase of the magnetic pressure than the thermal
pressure, as mentioned above. At a high enough density, the magnetic 
pressure will start to dominate the thermal pressure, forcing the inner
part of the cloud to flatten along the field lines (Tomisaka 1996). 
In the limit of a thin disk with frozen-in magnetic field, the field 
strength increases linearly with
the surface density, which is proportional to the square root of the 
mass density for an isothermal, self-gravitating disk in equilibrium 
along field lines (in the absence of appreciable magnetic 
squeezing; Ciolek \& Mouschovias 1994). Therefore, $B_0\propto 
\rho_0^{1/2}$ instead of $\rho_0^{2/3}$, and the magnetic pressure  will
not stiffen up the equation of state. In fact, in the presence of 
magnetic diffusion, the field strength will increase more slowly than
$\rho_0^{1/2}$, leading to a softening of the equation of state relative
to isothermal. This may explain the shallower-than-$r^{-2}$ slope of the 
density profile in the envelope found by Ciolek \& Mouschovias (1994;
see also Basu \& Mouschovias 1994). Furthermore, the magnetic tension
tends to dominate the magnetic pressure force in a disk geometry. It
effectively dilutes the gravity (Shu \& Li 1997), leading to a slower 
cloud contraction than the spherical case where the tension force is 
ignored on purpose (Safier et al. 1997). Ideally,
one should develop 2-D axisymmetric codes to treat realistic molecular
clouds with the magnetic forces comparable to the thermal force, 
perhaps along the line of Fiedler 
\& Mouschovias (1993) but with a weaker initial magnetic field and 
an improved treatment of the magnetic coupling. A further complication
that needs to be addressed in more realistic future calculations is the 
presence and decay of (magnetic) 
turbulent pressure, which may play an important 
(or even dominant) role in the core formation process (Nakano 1998; 
Myers \& Lazarian 1998; see also Lizano \& Shu 1989). 

\subsection{Dust Grains in Dense Molecular Cloud Cores} 

We have considered in our ``improved'' model one particular size distribution 
for dust grains - the standard MRN distribution. It enhances the magnetic
coupling considerably over the canonical value in an intermediate range
of densities. To gauge the effects of 
dust grains of other size distributions, one needs to recalculate the
number densities of all charged species. This has been done by Nishi et al. 
(1991) for several cases, and I shall rely exclusively on their results 
in the discussion below. 

As mentioned in \S~4.1, Nishi et al. (1991) calculated the number densities
of charged species for four models of grain size distributions: (1) the 
standard MRN distribution between $50\AA<a<2500\AA$; (2) the MRN 
distribution with ice mantles ($90\AA<a<4500\AA$); (3) the extended 
MRN distribution without ice mantles ($3\AA<a<2500\AA$), and (4) the 
standard MRN distribution plus an additional population of grains with 
a size of $a=4\AA$. For each of these four models, they computed
a magnetic dissipation timescale 
\begin{equation}
t_{_B}={3\over 4\pi G\rho^2}{A\over A_1},
\end{equation}
for an oblate cloud with a ``critical'' magnetic field strength $B_{cr}
\propto \rho^{1/2}$ and compared it to a free-fall timescale defined as
\begin{equation} 
t_f=\left({3\pi\over 32 G\rho}\right)^{1/2},
\end{equation}
which is somewhat different from our definition of freefall timescale
$t_{ff}=1/(4\pi G\rho)^{1/2}$ below equation (14). Both $t_{_B}$ and 
$t_f$ are plotted in their Figure~5 as a function of the number 
density of atomic hydrogen, $n_H$. It turns out that the ratio of these 
two timescales,
\begin{equation}
{t_{_B}\over t_f}={1.6\over (4\pi G)^{1/2}}{A\over \rho^{3/2}A_1},
\end{equation}
is almost identical to our definition of the magnetic coupling
coefficient $\nu_{ff}$ in equation (18). One can therefore 
roughly read off the coupling coefficient from their Figure~5, 
although the exact value depends on the field strength, which 
needs to be computed self-consistently in a dynamical model
such as ours. A comparison of
the timescale ratio in Nishi et al.'s Figure~5 for model (1),
with the standard MRN grain size distribution,  
and the coupling coefficient displayed in our Figure~4a shows
that they are indeed broadly similar, with a value close to
$10$ at low densities, and peaking somewhere between $10^7$ and
$10^8$cm$^{-3}$, before dropping off at still higher densities.
The timescale ratio for model (2), with ice-coated grains, shows
similar, although somewhat higher, enhancement at the intermediate
density range. We therefore expect the evolution of a magnetic
cloud with ice-coated grains to be similar to our improved case,
where dust grains of the standard MRN distribution is considered.
In particular, the characteristic, homologously collapsing core 
should also be present in spherical geometry, as is confirmed by 
a detailed calculation. 

The introduction of very small dust grains in models (3) and (4)
of Nishi et al. changes the magnetic coupling coefficient 
considerably. Naively, one expects that a large quantity small 
grains should make the coupling better. In reality, the ionization
level of the cloud is typically such that there is not enough 
charge for every small dust grain. As a result, most of the small
grains remain neutral, and are thus ``wasted'' as far as the magnetic 
coupling is concerned (see, however, Ciolek \& Mouschovias 1994
for a discussion about coupling between charged and neutral grains). 
From Figure~5 of Nishi et al., we find that the field dissipation 
timescale $t_{_B}$ remains about a factor of 10 above the free-fall 
timescale $t_f$ over a wide range of density in these two cases.
We therefore expect the magnetic coupling coefficient to be close 
to the canonical value $10$ and the evolution of clouds with these 
two model grain size distributions to resemble our reference 
model, as long as the density is below the decoupling density 
(which is of order $10^{10}$cm$^{-3}$ or higher).  

For dense molecular cloud cores, both theoretical studies and observations 
point to a grain-size distribution that could be substantially different 
from that in the diffuse interstellar medium (Evans 1996; Kr\"ugel \&
Siebenmorgen 1996): dust grains tend to acquire ice-mantles, coagulate and 
become fluffier. In particular, smallest grains ($\leq 50\AA$)
tend to be swept up quickly by large grains (Chokshi, Tielens \& 
Hollenbach 1993), while ice-coating can increase the size of 
grains by a factor up to two. These considerations lead to the conclusion 
that model (2) of Nishi et al. is perhaps the most plausible distribution 
in dense cores. For such a distribution, the magnetic coupling due to dust
grains should be enhanced above the value obtained in our improved 
model. Ideally, one would like to follow the change of 
grain-size distribution due to coagulation and ice-coating simultaneously
with the cloud dynamics. Such a task is beyond the scope of the present
paper. 

\section{Implications for Star Formation} 

\subsection{Structure and Kinematics of ``Starless'' Cores}

Our model core-formation calculations suggest that most of dense molecular 
cloud cores may be contracting at half a sound speed or more after the
central density increases by a factor of about $10^2$ from its initial 
value. This pre-collapse 
motion may help explain an interesting puzzle in some dense core properties.
Observations of N$_2$H$^+$ and C$_3$H$_2$ by Benson, Caselli \& Myers 
(1998) 
show that the two molecules, one is charged and one is not, move
more or less together, with a relative drift velocity between them 
smaller than $v_{_D}=0.03$ km s$^{-1}$. If the cores were nearly 
static and 
their evolution controlled by ambipolar diffusion, then for a typical
core radius of $0.05$ pc it will take at least 2 Myrs to allow
the central density to increase by a significant factor and form a
star. This would imply a typical lifetime for ``starless'' cores
an order of magnitude longer than that inferred from statistics
of dense N$_3$H cores associated with IRAS sources (Beichman et al. 1986),
which is a few times $10^5$ years (e.g., Shu 1995; Walmsley 1998). 
An obvious resolution of this discrepancy, as suggested by Figure 3d,
is that the ions and neutrals are contracting more
or less together, at a speed of, say $0.1$ km s$^{-1}$ or more, much 
larger than the ion-neutral drift speed (less than about 0.02 
km s$^{-1}$ in the improved model) once a dense core is formed out
of the initial, less dense background cloud. For an initial 
cloud central density of $10^3$cm$^{-3}$, our calculations suggest
that a ``starless'' dense core with a {\it central} density of 
$10^5$cm$^{-3}$ (the average density could be somewhat smaller; 
Mizuno et al. 1994) lasts about $2\times 10^5$ years before 
the central density formally goes to infinity and becomes a 
``starred'' core. Such a relatively short timescale is consistent 
with the lifetime of typical ``starless'' cores (Fuller \& Myers 
1987). 

One might argue that the substantial contraction motion in our model 
``starless'' cores may have something to do with the particular outer 
boundary conditions we used: a freely moving pressure boundary. I
have tried several other boundaries, including one with the outer edge 
of the cloud completely fixed in space. Even in such an extreme case
designed to minimize cloud contraction, most of the cloud matter 
away from the edge still infalls at a speed close to half an isothermal 
sound speed when the central density increases by a factor of $10^2$,
say from $10^3$ to $10^5$ cm$^{-3}$, as in our reference and improved 
models. Therefore, substantial inward motion appears to be characteristic
of dense ``starless'' cores formed out of roundish magnetized molecular 
clouds via ambipolar diffusion.  

Recent molecular line observations by Tafalla et al. (1998) (see also 
Myers \& Mardones 1997) and Williams et al. (1999) seem to indicate 
substantial inward motion of both ions (N$_2$H$^+$) and neutrals (CS) 
with velocities up to 0.1 km s$^{-1}$ in the ``starless'' core 
L1544. An interesting feature of the inward motion, as noted by
Williams et al. (1999), is that the maximum infall speed is 
essentially the same on both the small ($\sim 0.01$ pc) and the
large ($\sim 0.1$ pc) scales. Within the context of our spherical 
models, such inward motion with velocities comparable to half an 
isothermal sound speed on both size-scales occurs when the central 
cloud density enhancement is between roughly $10^2$ and $10^3$; beyond 
$10^3$, the maximum inward velocity becomes comparable to, or even 
larger than, the isothermal sound speed. Similar conclusions can be 
reached for disk-like configurations, as long as the initial 
mass-to-flux ratio is too far from the critical value (Crutcher 
1999) and the clouds are not too centrally concentrated to
begin with (Li 1999). 

Another possible example of ``starless" cores with ``infall" 
signatures is L1498 (Lemme et al. 1995), which shows a pronounced 
double-peaked CS feature with the blue peak stronger than the red 
peak. It is perhaps surprising that not many more such sources 
are found, especially among more advanced cores (i.e., those with 
higher central densities), such as those studied in millimeter 
dust continuum (e.g., Andr{\'e}, Ward-Thompson \& Motte 1996). 
It seems that a substantial motion is required of ``starless'' 
cores by their typical size and lifetime, independent of theoretical 
models: to shrink a dense core of about 0.05 pc in radius by half 
(so that the average density would increase by an order of 
magnitude) in $2\times 10^5$ years demands a contraction speed 
of $0.1$ km s$^{-1}$ - about half of the isothermal sound speed 
of a typical 10 K cloud core. One possibility is that the expected 
pre-collapse inward motion shows up in the non-thermal ``turbulent'' 
component of the velocity dispersion which, interestingly, is also
about half of the thermal component for many ``starless'' NH$_3$ 
cores (Fuller \& Myers 1987; Barranco \& Goodman 1998). 

Besides inward motion, another property of dense cores that could 
be deduced in principle from a variety of observations is the density 
profile. It is now established that ``starless'' cores have a
plateau-envelope type profile (with a more-or-less constant
density plateau at the center surrounded by an envelope whose
density decreases with radius roughly as a power-law; Andr{\'e}, 
Ward-Thompson \& Motte 1996) while the plateau region is absent in 
``starred'' cores, in accordance with theoretical expectations (e.g., 
Lizano \& Shu 1989; Basu \& Mouschovias 1994; Ciolek \& Mouschovias
1994; Safier et al. 1997; Figures~1a and 3a of this paper; and
Shu 1977). Observational estimates of the power-index in the envelope
in the literature vary from shallower than $-1$ to steeper than
$-2$. Theoretical models of Ciolek \& Mouschovias (1994) based
on thin-disk approximation predict a power index between about 
$-1.5$ to $-1.85$ for clouds with an initial magnetic pressure 
much higher than the thermal pressure. It is substantially shallower 
than that of an singular 
isothermal sphere ($-2$). It may not be applicable to those cloud cores
that have an initial magnetic pressure comparable to, or smaller
than, the thermal pressure and thus are more roundish. In such cases, 
the calculations presented here 
suggest that the density profile in the envelope could be 
considerably steeper. Indeed, taken at the face value, Figures~1c
and 3c yield a power index close to $-2.3$ in our idealized 
spherical models. Inclusion of magnetic tension may flatten 
the density profile somewhat, although it is not clear whether 
it can become as flat as $-2$. It is interesting to note in this
regard the 2-D calculations of Tomisaka (1996), who considered 
the collapse of relatively weakly magnetized axisymmetric cloud
with frozen-in field. The magnetic tension is taken into account
properly. He find a density profile in the envelope ranging
from $r^{-2.1}$ (his Model A) to $r^{-2.4}$ (his Model B) in the 
equatorial plane of the cloud. Future 2-D calculations along similar 
lines but with proper treatment of the magnetic coupling should 
be able to settle this issue.

Observationally, there are some indications that the density profile 
in the envelope is as steep as $r^{-2}$, maybe even steeper. In the 
best studied case, L1689B, using millimeter dust continuum, Andr{\'e}
et al. (1996) conclude that the power-index of the density profile 
is consistent with $r^{-2}$ (perhaps even slightly steeper; see their
Figure~3b). An extreme example of steep density profile is given by
Kane \& Clemens (1997), whose deduced a median apparent volume density 
profile for a sample of ``starless'' Bok globules to be as steep as 
$r^{-2.6}$ from molecular line observations. Similarly, Abergel et
al. (1996) concludes from ISO absorption observations that the 
``starless" dense core, B2, in $\rho$ Oph may have an envelope 
whose density drops as fast as $r^{-3}$. In any case, through
concrete numerical examples, we have demonstrated the possibility 
that magnetic pressure may stiffen the equation of state of a
magnetic cloud beyond isothermality (i.e., making the effective
adiabatic index $\gamma\ > \ 1$). Collapse of polytropic gas spheres 
with $\gamma >1$ will also lead to a plateau-envelope density profile, 
with the density power index in the envelope approaching $2/
(\gamma-2)$ (Yahil 1983). Therefore, stiffening of the equation 
of state beyond isothermality tends to produce an envelope density
profile steeper than $r^{-2}$. It is not clear as to what extent 
such a tendency is geometry-dependent. 

\subsection{Protostellar Mass Growth}

An important quantity in the star formation theory is the rate of mass 
accretion onto the forming protostar. On dimensional grounds, it
should be proportional to $a^3/G$ (where $a$ is the isothermal 
sound speed and $G$ the gravitational
constant; Shu 1977), although the
value of the proportionality constant is currently under debate. 
As a starting point for discussion, we consider the 
well known ``inside-out" collapse model of static, nonmagnetized 
singular
isothermal sphere (Shu 1977). The mass accretion rate onto the
central compact object is found by self-similarity technique to
be 
\begin{equation}
{\dot M}=0.975 {a^3\over G}.
\end{equation}
For a 10 K gas typical of Taurus star formation region, 
$a=0.2$ km s$^{-1}$, so that ${\dot M}=2\times 10^{-6}$M$_
\odot$yr$^{-1}$. It corresponds to an accretion luminosity of 
\begin{equation}
L_{acc}={GM{\dot M}\over R}=10\ \ L_\odot \left({M\over 0.5 M_\odot}
	\right)\left({a\over 0.2\ \ \hbox{\rm km/s}}\right)^3\left({3 
	R_\odot\over R}\right),
\end{equation}
where $M$ and $R$ are the mass and radius of the central protostar, 
assuming all of ${\dot M}$ lands directly on the stellar surface. 
Therefore, for a typical low mass star of $0.5$ M$_\odot$ and 
3 R$_\odot$, the accretion luminosity predicted by this simple
model is about 10 L$_\odot$ (cf, Shu 1995). Compared with the typical
observed luminosity of about 4 L$_\odot$ by Myers et al. (1987; see
also Gregersen et al. 1997), 
it is too large by only a modest factor of a few. The apparent 
discrepancy between the model prediction of accretion luminosity 
and the observed luminosity is sometimes referred to as the 
``luminosity problem" in the star formation literature. The problem 
is exacerbated by the presence of magnetic fields envisaged in 
the current paradigm. 

Magnetic fields could enhance mass accretion rate in three ways. First, 
compared with nonmagnetic cases, a magnetized cloud core can 
have a higher initial equilibrium density afforded by the additional 
magnetic support. Collapse of a denser core leads naturally to a 
higher mass accretion rate (Li \& Shu 1997). Second, magnetic fields 
increase signal speed, leading to a larger collapsing region at
any given time, although this increase tends to be cancelled to
a large extent by magnetic tension, which retards the collapsing
flow (Galli \& Shu 1993). 
The last, perhaps also the most subtle, effect of magnetic fields
is to stiffen the equation of state of the cloud (at least in
a spherical geometry), leading to a
substantial steepening of density profile in the envelope and an
increase in the cloud contraction speed. The net result is the presence
of a large initial peak in the mass accretion rate (Safier et al.
1997; Li 1998), also evident from some of nonmagnetic 
(Foster \& Chevalier 1993; Henriksen, Andr{\'e} \& Bontemps 1997)
as well as field-frozen (Tomisaka 1996) collapse calculations. This
effect is illustrated most clearly by the our improved model.  
 
The enhancement of magnetic coupling by dust grains yields a more or 
less homologously collapsing central region of substantial mass. At
the end of the cloud evolution in our improved model (when the central 
number density reaches a value of $10^{11}$ cm$^{-3}$) a region of 
about $0.62$ M$_\odot$ is collapsing toward the center with a speed of 
$1.55$ km s$^{-1}$ near its edge at a radius $9.85\times 10^{-4}$ pc. 
Unless slowed down somehow by a strong force, such as the centrifugal force 
associated with rotation, the collapse tends to accelerate and all of 
this mass will reach the origin in less than $620$ years. It would 
imply a stunning average (for the first $0.62$ M$_\odot$) mass 
accretion rate of more than $10^{-3}$ M$_\odot$ yr$^{-1}$! This 
would produce an enormous accretion luminosity (of order $10^4$ 
L$_\odot$ from equation [24]) if most of the accreted mass is
channeled to the stellar surface quickly. Although there are indirect 
observational indications 
that the mass accretion rate may be as much as ten times higher
in the ``Class 0'' phase than in the ``Class I'' phase (see 
Andr{\'e} 1997 for a review), ways must be found to 
reduce the above estimate by two to three orders of magnitude. 
A proper inclusion of magnetic
tension will undoubtedly reduce the mass accretion rate somewhat, 
although the extent of such a reduction is unclear at present. This
potentially serious ``luminosity problem'' for spherical models 
makes two dimensional treatments of the cloud evolution with 
appropriate magnetic coupling all the more pressing. 

Another way of reducing the mass accretion rate is through outflows. 
Powerful molecular outflows are generated early on in the protostellar 
mass growth process. They are observed in most of the youngest, the
so-called ``Class 0'', sources (e.g., Barsony 1994) and many older,
``Class I'' sources, and should play a role in reversing the mass 
accretion (Shu et al. 1987; Velusamy \& Langer 1998).  
Otherwise, mass in the envelope will keep falling towards the center,
building up a star much more massive than the half solar mass of
a typical low-mass star. The accretion reversal process could be
extremely efficient. Indeed, observations of L1551 IRS 5 indicate that
the mass outflow rate on a 
sizescale of order 2000 AU is inferred to be comparable to, or
even larger than, the mass infall rate 
(Fuller 1994). Details of the infall-outflow interaction remains to 
be worked out. 

A third possibility is to store the accreted mass in a circumstellar disk. 
The stored mass may be gradually fed onto the 
central star, perhaps mainly in short bursts as envisioned in Hartmann 
et al. (1993). However, it is unlikely for the circumstellar disk mass 
to exceed the stellar mass much, because of gravitational instabilities.
Alternatively, most of the accretion luminosity in the
disk could be released in a nonradiative form, spent perhaps in 
driving powerful outflows (Bontemps et al. 1996). Either way, the 
implications for the evolution of protostellar disks in general, 
and the proto-solar nebula in particular, would be profound. 

\section{Summary}

As a step towards a quantitative understanding of the star formation
process, I have followed numerically the evolution of magnetized, 
spherical clouds whose magnetic pressure is initially equal to the 
thermal pressure throughout the cloud. To gauge the dynamical 
effects of dust grains, I considered two models for the magnetic 
coupling coefficient, which controls the dynamical evolution of 
the cloud. In one model, the coupling coefficient is specified at 
a canonical value of $10$. In the other, it is evaluated 
self-consistently from ionization equilibrium, taking into account of 
dust grains with a standard MRN size distribution. My main conclusions 
are: 

(1) Substantial pre-collapse inward motion is expected of dense cores 
formed out of roundish magnetic clouds due to ambipolar diffusion. 
In our specific models, an inward speed of order half an isothermal 
sound speed is achieved
over much of the (10 M$_\odot$) cloud when the central density 
increases by a factor of $10^2$ above its initial value,
say from $10^3$ to $10^5$cm$^{-3}$. Such inward motion may have 
been observed 
recently in the ``starless'' core L1544 (Taffala et al. 1998;
Williams et al. 1999). It may
explain why typical ``starless'' cores 
last only for a few times $10^5$ years (Beichman et al. 1986; Shu 
1995), not much longer than their dynamical timescale. The 
subsonic pre-collapse motion could also 
be the origin of the small but significant nonthermal component of 
linewidths observed in many ``starless'' cores. 

(2) As found previously by several other authors, ``starless'' cores 
formed out of magnetic clouds always have a flat central plateau 
surrounded by an envelope whose 
density decreases with radius roughly as a power-law. Our model 
calculations suggest that the slope in the envelope is significantly 
steeper than those obtained by Mouschovias and collaborators,
who considered disk-like, rather than sphere-like, clouds with an initial 
magnetic pressure much higher than the thermal pressure. 

(3) Dust grains of a standard MRN size distribution can enhance 
the magnetic coupling coefficient above the canonical value of 
10 considerably (Nishi et al. 1991). The enhancement peaks 
somewhere between
$10^7$-$10^8$ cm$^{-3}$, with a maximum value about an order of
magnitude higher than the canonical value. The enhanced coupling 
allows the cloud matter to trap the magnetic flux better during 
the advanced, more dynamic phase of the core formation. In the 
idealized spherical geometry, the trapping of
flux creates a central region of substantial mass that collapses almost
homologously. In our particular example, more than half a solar mass 
can reach
the center in less than $10^3$ years, greatly exacerbating the
accretion ``luminosity problem'' of star formation. Magnetic 
tension may alleviate this problem. Two dimensional, fully dynamic 
calculations are urgently needed. 

\acknowledgements 
I thank Frank Shu, R. Nishi and the referee, Steve Stahler, for useful 
comments and R. Nishi for an ionization code.

\newpage

\includegraphics[scale=0.8]{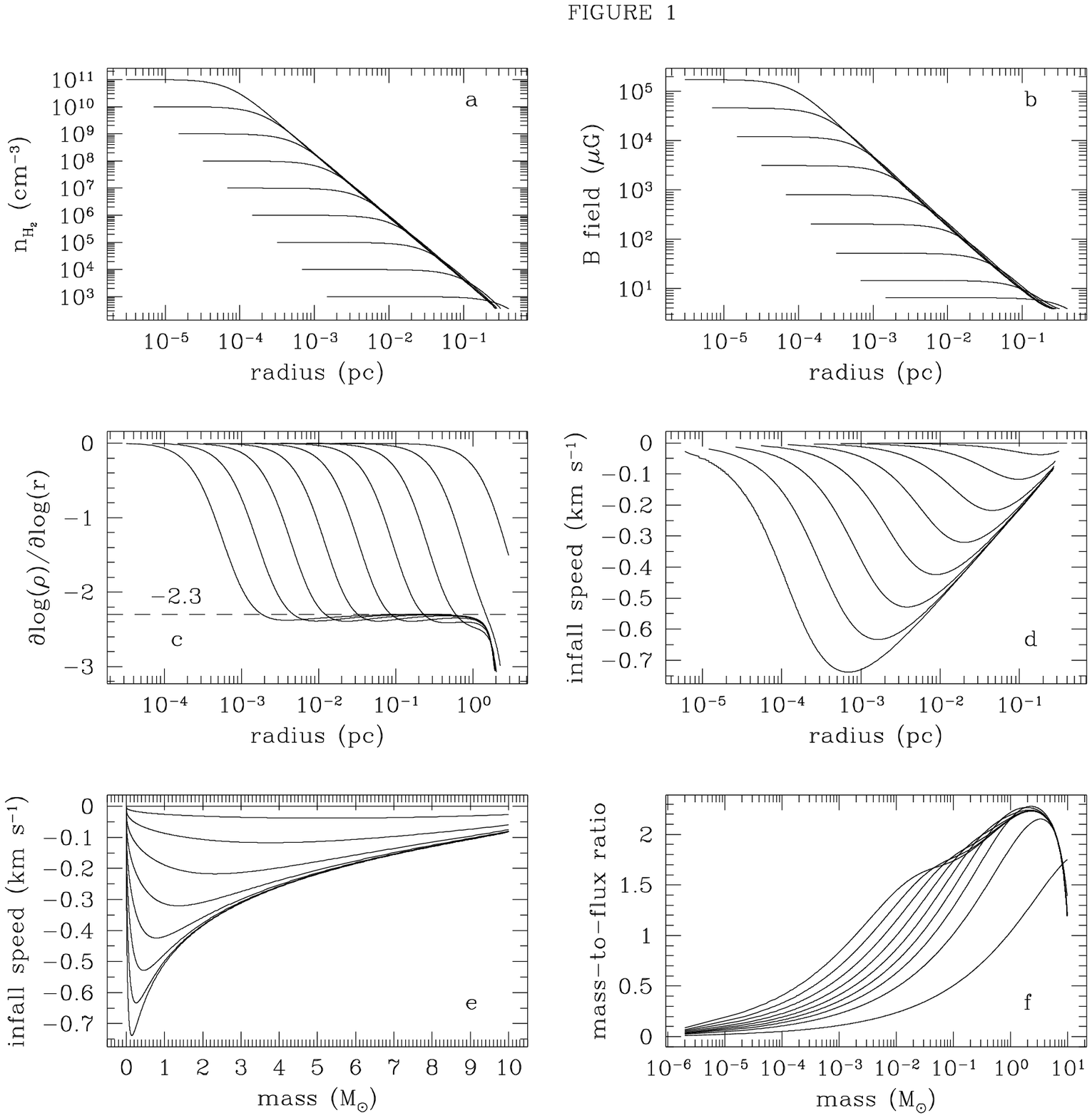}
\hskip 1cm

\figcaption[fig1.ps]{The distributions of various cloud quantities for 
the reference
model with a constant magnetic coupling coefficient of $10$. In each 
panel, we plot the initial distributions and the distributions when the
central density increases by eight successive factors of $10$. Shown
in the panels are (a) the number density of molecular hydrogen; (b) the
strength of the magnetic field; (c) the log-log slope of density profile;
(d) the infall speed as a function of radius; (e) the infall speed as a
function of mass; and (f) the mass-to-flux ratio in units of the critical
ratio $1/(2\pi G^{1/2})$.
\label{reference}}

\newpage

\includegraphics[scale=0.8]{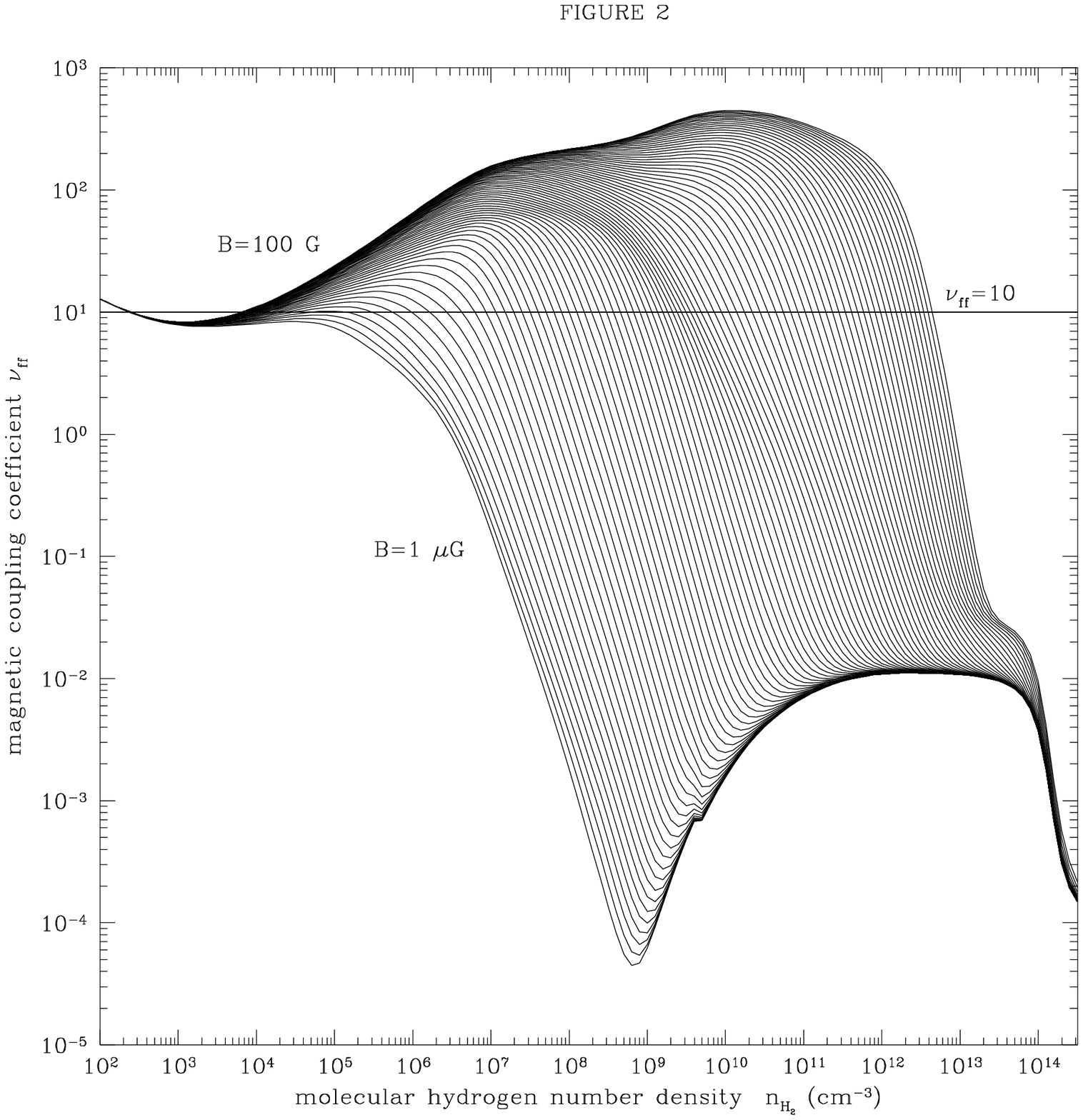}
\hskip 1cm

\figcaption[fig2.ps]{Each curve represents a distribution of the 
magnetic coupling 
coefficient $\nu_{ff}$ as a function of density for a given field 
strength. The field strength increases by 80 successive factors of $10^{0.1}$ 
from a minimum value of $1\ \mu$G to a maximum value of $100$ Gauss. 
Dust grains of a standard MRN size distribution are
taken into account and the coefficient $\nu_{ff}$ shown here is used in the 
numerical calculation of the improved model.
\label{coupling}}    

\newpage

\includegraphics[scale=0.8]{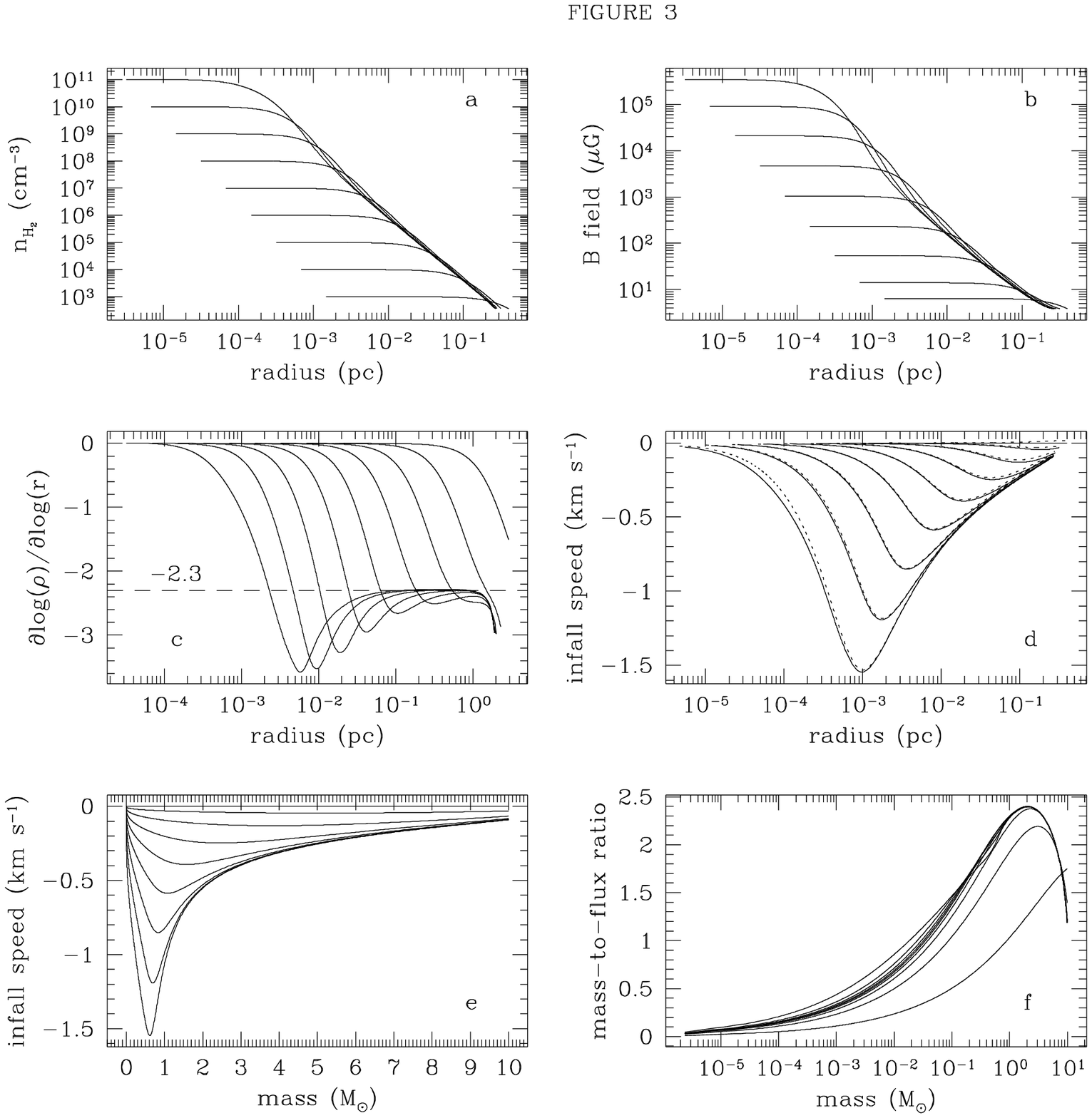}
\hskip 1cm

\figcaption[fig3.ps]{The distributions of various cloud quantities 
for the improved
model where the magnetic coupling coefficient is computed 
self-consistently from ionization equilibrium. In each 
panel, we plot the initial distributions and the distributions when the
central density increases by eight successive factors of $10$. Shown
in the panels are (a) the number density of molecular hydrogen; (b) the
strength of the magnetic field; (c) the log-log slope of density profile;
(d) the infall speeds of neutrals (solid lines) and molecular ions
(dotted lines) as a function of radius; (e) the infall speed as a
function of mass; and (f) the mass-to-flux ratio in units of the critical
ratio $1/(2\pi G^{1/2})$. Compare these panels with those in Figure~1
to gauge the dynamical effects of dust grains.
\label{improved}} 

\newpage

\includegraphics[scale=0.8]{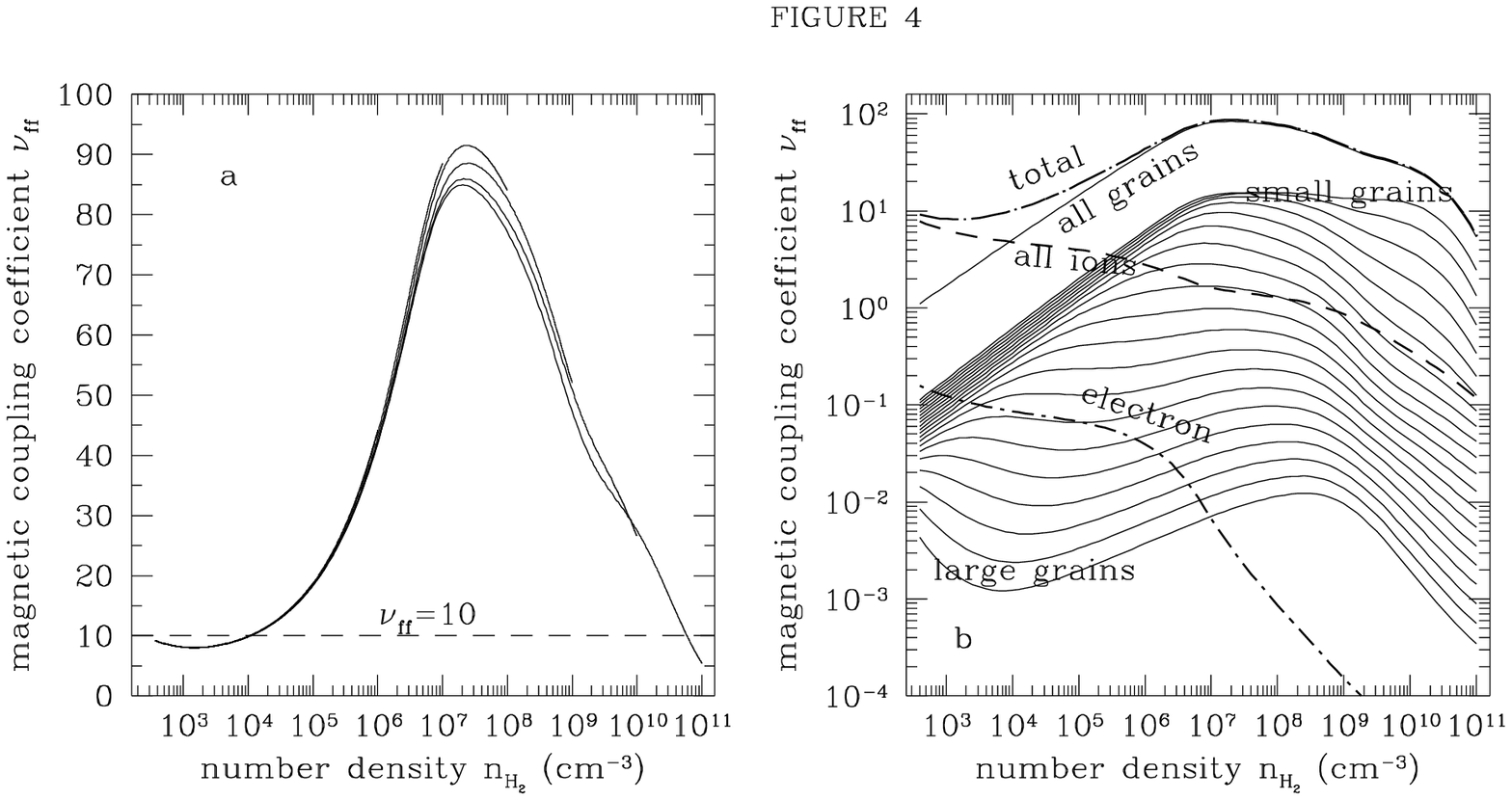}
\hskip 1cm

\figcaption[fig4.ps]{(a) plotted are the self-consistently 
calculated magnetic coupling 
coefficient $\nu_{ff}$ as a function of cloud density in the improved model. 
As in Figure~3, the initial distribution and  the distributions when the
central density increases by eight successive factors of $10$ are shown; 
(b) the contributions from various charged species to the total magnetic
coupling coefficient at the end of the run when the central cloud density
reaches $10^{11}$ cm$^{-3}$. Note that grains dominate ions at densities 
above $10^4$ cm$^{-3}$, and small grains contribute more than large
grains, as expected.  
\label{species}}

\end{document}